\documentclass{iau_FM}
\usepackage{graphicx}
\usepackage{natbib}

\def\hho{H$_2$O}
\def\ohp{OH$^+$}
\def\hhop{H$_2$O$^+$}
\def\hhoe{H$_2^{18}$O}
\def\hhhop{H$_3$O$^+$}

\def\ccm{cm$^{-3}$}
\def\mic{$\mu$m}
\def\gtsim{{_>\atop{^\sim}}}

\def\rs{s$^{-1}$}

\title[FM 15.~~Water and organics in the Universe] 
{Water in the interstellar media of galaxies}

\author[Floris van der Tak]   
{Floris F.S. van der Tak}

\affiliation{SRON \& Kapteyn Astronomical Institute, P.O. Box 900,
9700 AV Groningen, the Netherlands \\ email: {\tt vdtak@sron.nl}}

\pubyear{2015}
\jname{Astronomy in Focus, Volume 15} 
\editors{Kwok \& Bergin, eds.}
\begin{document}

\maketitle

\begin{abstract} 
This paper reviews recent observations of water in Galactic interstellar clouds and 
nearby galactic nuclei. Two results are highlighted:
(1) Multi-line \hho\ mapping of the Orion Bar shows that the water chemistry in PDRs is driven by photodissociation and -desorption, unlike in star-forming regions.
(2) High-resolution spectra of \hho\ and its ions toward 5 starburst / AGN systems reveal low ionization rates, unlike as found from higher-excitation lines.
We conclude that the chemistry of water strongly depends on radiation environment, and that the ionization rates of interstellar clouds decrease by at least 10 between galactic nuclei and disks.
\end{abstract}

\firstsection 
\section{Introduction}

The stars in galaxies are part of a cycle of matter: from diffuse clouds to dense star-forming regions and protoplanetary disks.
The water molecule is a great tool to understand the conditions of interstellar matter along this cycle, which can be used in 2 different ways.
First, water acts as a natural chemical filter, because its abundance varies strongly with the temperature. At low dust temperatures ($<$100\,K), water freezes out on grain surfaces, while at high gas temperatures ($>$250\,K), \hho\ abundances are enhanced by neutral-neutral reactions. Thus, water acts as a probe of energy injection into the medium, as opposed to species like CO which measure the reservoir of matter.
Second, water is a sensitive tracer of physical conditions. Its asymmetric structure leads to a rich line spectrum covering large ranges in radiative lifetimes, so that \hho\ line ratios are sensitive to both gas density and kinetic temperature. Being a hydride, its low reduced mass shifts its rotational line spectrum to high ($\sim$THz) frequencies, so that the lines are also probes of dust continuum radiation.
See \citet{ewine2013} for a review of interstellar \hho\ physics and chemistry.

Interstellar water is well known from its maser lines at radio frequencies, which are useful to probe circumstellar kinematics \citep{sanna2012} and also to measure accurate distances throughout the Galaxy \citep{reid2014}. 
Recently, thermal \hho\ lines have also been used to measure the distance to an interstellar cloud \citep{giannetti2015}.
Using water to probe physical and chemical conditions depends on measuring thermal \hho\ lines, which almost without exception require observation from space.
Compared with previous missions (ISO, Spitzer, SWAS, Odin), ESA's {\it Herschel} space observatory \citep{pilbratt2010} offers great advances in spatial and spectral resolution and sensitivity to \hho\ rotational lines spanning excitation energies of 0--1000\,K and critical densities of 10$^6$--10$^9$\,\ccm.
While HIFI offers high spectral resolution for low-$J$ lines, PACS offers sensitive mapping for higher-$J$ lines.

This paper describes recent Herschel results on \hho\ in the large-scale interstellar medium.
For the other parts of the interstellar matter cycle, we refer to the contributions to this volume by Caselli (pre-stellar cores), Neufeld \& J{\o}rgensen (protostars), and Pontoppidan (protoplanetary disks).

\begin{figure}[b]
\begin{center}
\includegraphics[width=7cm,angle=0]{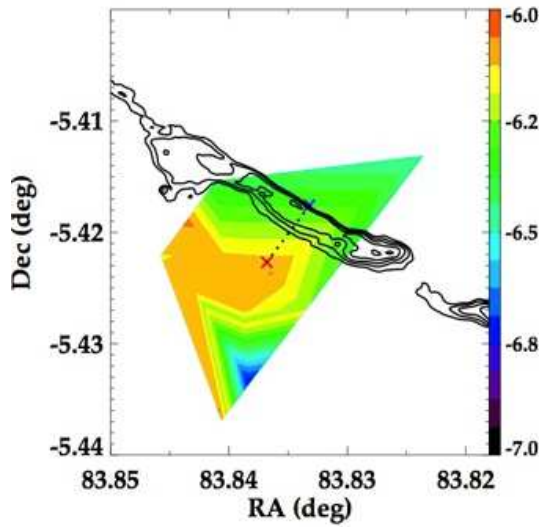} 
\caption{Map of the \hho\ abundance in the Orion Bar, with contours of Spitzer 8~\mic\ overlaid. The red cross denotes the \hho\ abundance peak and the blue cross the position of the ionization front. From: Choi, Bergin \& van der Tak (in prep).}
\label{fig:o-bar}
\end{center}
\end{figure}

\begin{figure}[b]
\begin{center}
\includegraphics[width=5cm,angle=0]{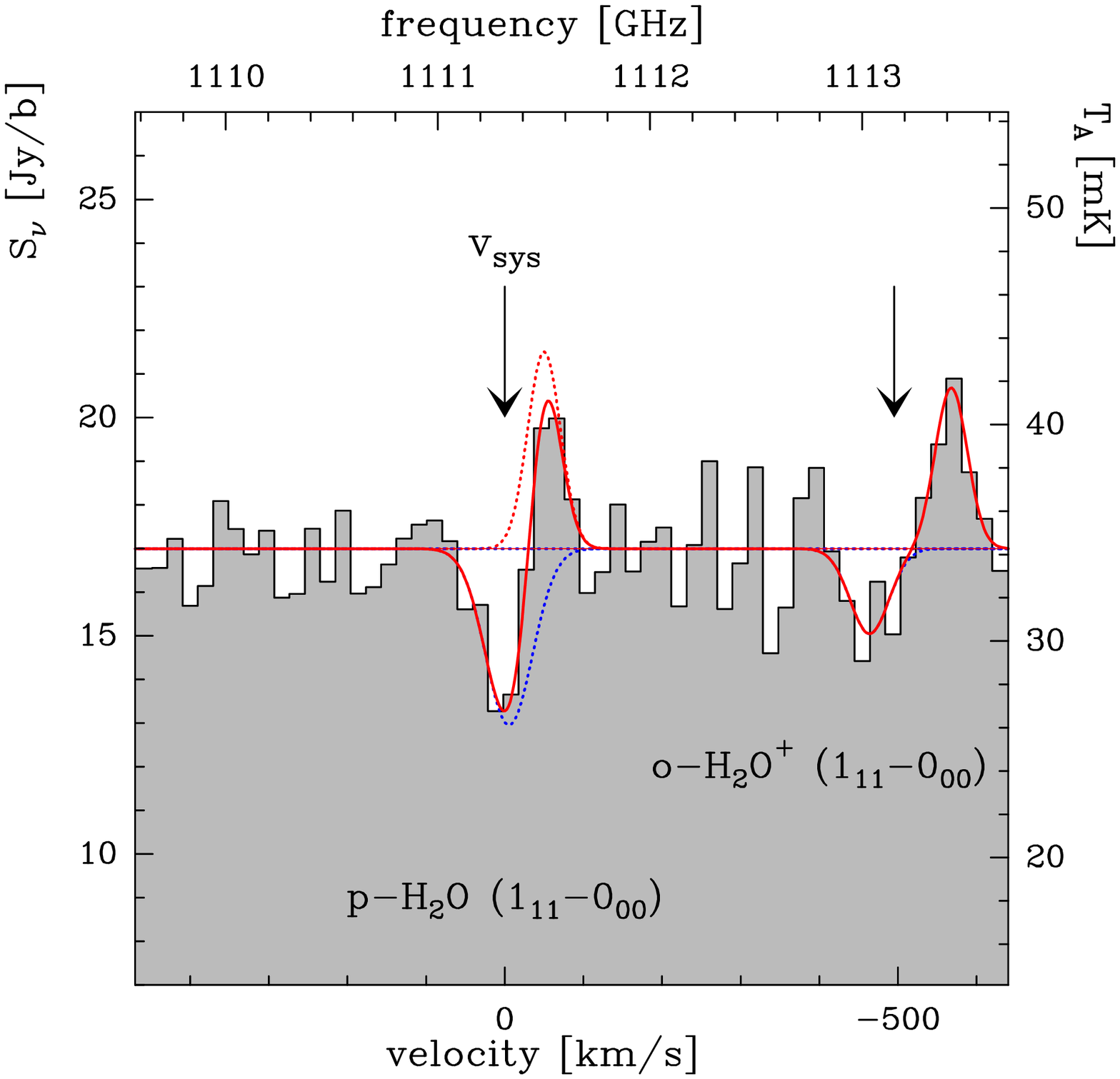} 
\includegraphics[width=5cm,angle=0]{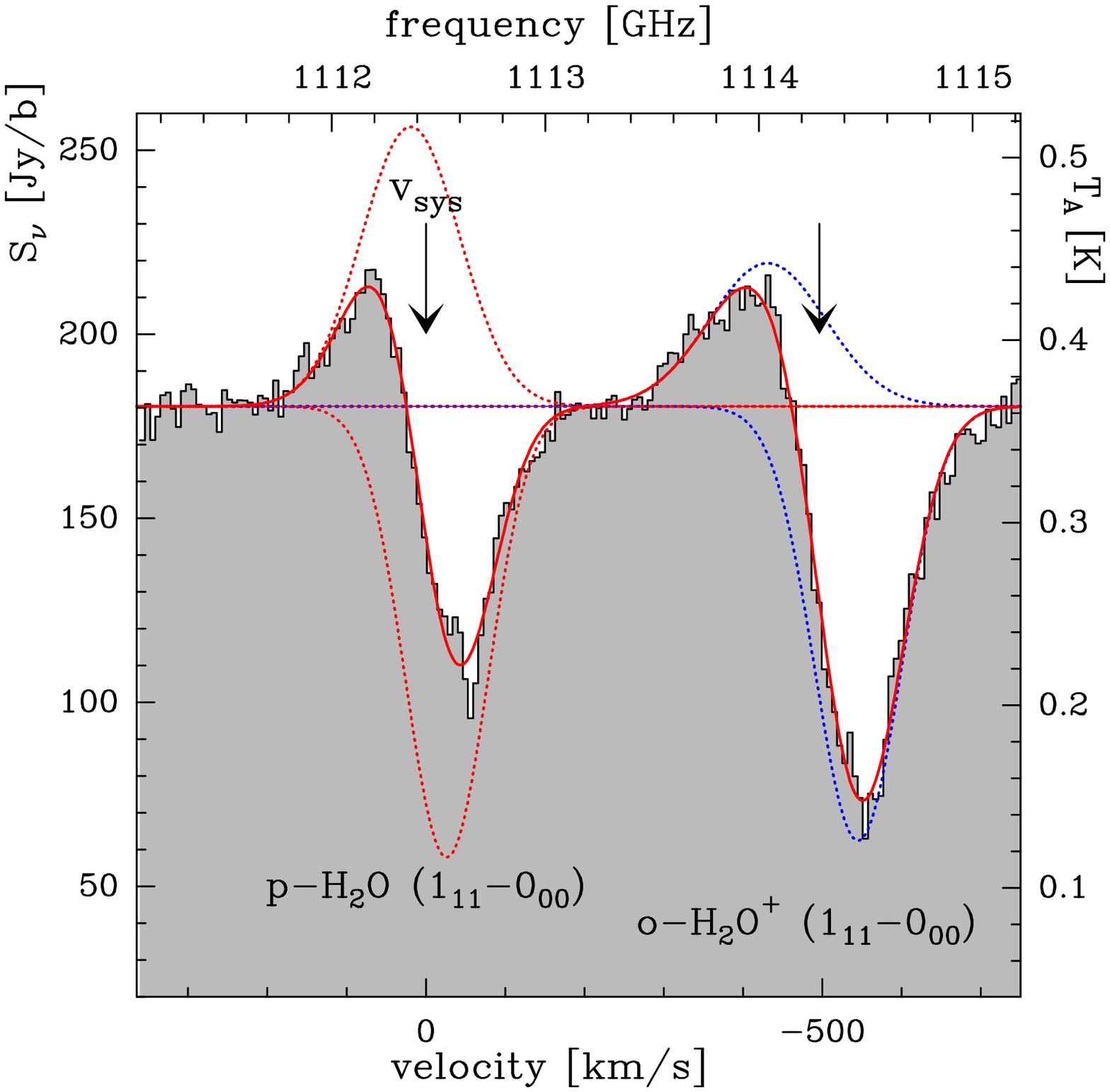} 
\caption{Spectra of the \hho\ and \hhop\ ground-state lines toward Cen A and NGC 253, showing evidence for infall and outflow motions. From: Van der Tak, Weiss, Liu \& G\"usten (in prep).}
\label{fig:hexgal}
\end{center}
\end{figure}

\section{Water in photon-dominated regions (PDRs)}

One field where Herschel has allowed significant progress are the physics and chemistry of PDRs, which are places where the surfaces of dense interstellar clouds are irradiated by nearby hot (OB-type) stars.
For homogeneous clouds, standard calculations \citep{hollenbach1997} predict a layered structure, which is indeed observed \citep{vdwiel2009} although additional clumpy structure is also seen \citep{lis2003}.

The roles of grains and of radiation in driving the oxygen chemistry in PDRs is unclear, but a model by \citet{hollenbach2009} predicts that radiation regulates the \hho\ abundance.
Deep inside the cloud, \hho\ is predicted to freeze out onto dust grains, while close to the surface, \hho\ is photodissociated by impinging UV radiation. 
The \hho\ abundance peaks at intermediate depths ($A_V$=3--8 mag), due to photodesorption of the icy grain mantles.
In this model, \hho\ traces the surface of the cloud, which a correlation study of \hho\ with various other tracers based on large-scale maps of the Orion molecular cloud seems to confirm \citep{melnick2011}.

To further test the Hollenbach model, Choi, Bergin \& van der Tak (in prep) have used HIFI maps of 7 low-$J$ lines of \hho\ toward the prototypical nearby Orion Bar PDR (Fig.~1). 
The \hho\ emission is seen to trace the Bar, and to peak between the C$^{18}$O emission, which traces the cloud interior, and the C$_2$H emission, which traces the cloud surface.
The authors use a non-LTE model \citep{vdtak2007} to estimate the \hho\ column density for each map position, and combine this with the C$^{18}$O emission to estimate the \hho\ abundance distribution in the Orion Bar. 
The abundance peaks at an offset of $\approx$22$''$ from the ionization front as traced by the edge of the Spitzer 8\,\mic\ emission in Fig.~1. 
Assuming a distance of 420 pc \citep{menten2007} and an average gas density of 10$^5$\,\ccm, this offset corresponds to $A_V$=8 mag, which is at the high end of the range predicted by \citet{hollenbach2009}. 
We regard this result as a confirmation of Hollenbach's model, and an indication of the importance of photodesorption in PDRs.

Further evidence for photodesorption processes in the Orion PDR comes from the subthermal ortho/para ratio of \hho\ \citep{choi2014}. 
Non-LTE modeling of the ortho- and para-\hhoe\ ground state lines observed with HIFI indicate a spin temperature of 10--20\,K, which is well below the temperatures of dust and gas, ruling out thermal ice evaporation and gas-phase reactions as mechanisms for \hho\ production. 
Photodesorption of icy grain mantles may be a possible mechanism, although it is uncertain if the spin temperature of solid \hho\ is preserved upon photodesorption.

Low ortho/para ratios for \hho\ are also observed for clouds along the line of sight toward the Galactic center \citep{lis2013}, although the derived spin temperatures of 24--32\,K are higher than in Orion so the case for photodesorption is less clear.
Finally, cosmic-ray induced photodesorption is likely at work in the centers of dense pre-stellar cores, as indicated by the surprisingly high abundances of \hho\ and organic species \citep{caselli2012,bacmann2012}.

\section{Water in galactic nuclei}

The far-infrared lines of water are also powerful tracers of the physical conditions in the gas in galactic nuclei. 
Spectral scans with Herschel/PACS have revealed numerous \hho\ lines towards starbursts and AGN, which reveal both compact warm nuclear gas and an extended cold disk component which dominates by mass. 
The abundances of \hho, HCN and OH are enhanced in the nuclear gas, probably due to grain mantle evaporation and/or cosmic-ray or X-ray irradiation \citep{eduardo2012}.

Models by \citet{eduardo2014} show that while collisional excitation dominates for low-$J$ lines, radiative pumping dominates for $E_u \gtsim 250$\,K, which explains why the luminosities of the $J$=0--2 lines decreases with dust temperature, while higher-$J$ lines show the opposite behaviour.

When combined with its associated ions, water is also useful as a probe of the ionization rates of galactic nuclei. 
Van der Tak et al (in prep) have used Herschel/HIFI to observe the ground-state lines of \hho, \hhop\ and \ohp\ toward 5 nearby starbursts and AGN. 
The line profiles range from pure absorption to P~Cygni indicating outflow and inverse P~Cygni indicating infall motions (Fig\,\ref{fig:hexgal}).
The \hho/\hhop\ ratios of a few indicate an origin of the lines in diffuse gas, and the \ohp/\hhop\ ratios suggest a molecular fraction of $\approx$11\% for the gas. 
However, the low \hho\ abundance may indicate enhanced photodissociation by UV from the nuclei or depletion of \hho\ onto dust grains.

Adopting recent Galactic values for the average cloud density and the ionization efficiency, Van der Tak et al (in prep) estimate cosmic-ray ionization rates of
$\zeta_{\rm CR}$ $\sim$10$^{-16}$ \rs, similar to the value for the Galactic disk, but somewhat below that of the Galactic center and well below that of AGN estimates from excited-state \hhhop\ lines \citep{goto2014}. 
They conclude that the ground-state lines of \hho\ and \hhop\ probe primarily non-nuclear gas in the disks of these centrally active galaxies.
Their data thus provide evidence for a decrease in ionization rate from the nuclei to the disks of external galaxies, as found for the Milky Way \citep{indriolo2015}.


\section{Conclusions and future outlook}

From the above studies, we conclude that the chemistry of interstellar water strongly depends on environment.
The bulk of \hho\ is formed in dense clouds on the surfaces of cold dust grains, while in hot shocked gas, gas-phase production in neutral-neutral reactions dominates.
The fresh water gets into the gas phase by thermal desorption in protostellar envelopes, but by photodesorption in PDRs and pre-stellar cores.
Conversely, water is removed form the gas phase by freeze-out in cold dense clouds and protostellar envelopes, but by photodissociation in diffuse clouds and PDRs.
Third, the excitation of \hho\ depends on the line: low-$J$ lines tend to be excited by collisions, while high-$J$ lines are pumped by dust continuum radiation.
Finally, \hho\ and its ions can be used to show that the ionization rate of gas in galaxies varies with environment: from 10$^{-17}$--10$^{-16}$\,\rs\ in galactic disks to 10$^{-15}$--10$^{-14}$\,\rs\ in galactic nuclei.

Starting in 2016, ALMA Band 5 will offer (sub)mm observations of thermal \hho\ at high angular and spectral resolution, especially warm \hho\ such as found in the snowline regions of protoplanetary disks.
The mid-infrared (MIRI) instrument onboard JWST, to be launched 2018, will offer high sensitivity to hot \hho, for instance in the planet-forming regions of circumstellar disks.
In the late 2020's, the SPICA space telescope will offer mid- and far-infrared spectroscopy at high enough sensitivity to probe warm \hho\ in galactic nuclei and disks, out to the early Universe, where ground-based \hho\ observations show great promise \citep{omont2013}.
Clearly the study of interstellar water has just begun!

\bibliographystyle{aa}
\bibliography{vdtak-fm15}






\end{document}